\documentclass[12pt]{article}
\usepackage{graphicx}
\usepackage[compress]{cite}
\usepackage{caption}
\newcommand\pubnumber{DPF2015-99}
\newcommand\pubdate{October 25, 2015}

\providecommand{\e}[1]{\ensuremath{\times 10^{#1}}}

\def\uva{University of Virginia, Charlottesville VA, USA}
\def\fnal{Fermi National Accelerator Laboratory, Batavia IL, USA}

\def\Title#1{\begin{center} {\Large #1 } \end{center}}
\def\Author#1{\begin{center}{ \sc #1} \end{center}}
\def\Address#1{\begin{center}{\scriptsize \it #1} \end{center}}
\def\mailto#1{{\tt #1}}
\def\ead#1{\vspace*{5pt}\Address{E-mail: \mailto{#1}}}
\newcommand\pubblock{\rightline{\begin{tabular}{l} \pubnumber\\
         \pubdate  \end{tabular}}}
\newenvironment{Abstract}{\begin{quotation}  }{\end{quotation}}
\newenvironment{Presented}{\begin{quotation} \begin{center} 
      \begin{large}}{\end{large}\end{center} \end{quotation}}
\def\Acknowledgments{\bigskip  \bigskip \begin{center} \begin{large}
             \bf ACKNOWLEDGMENTS \end{large}\end{center}}

\def\beq{\begin{equation}}
\def\eeq#1{\label{#1}\end{equation}}
\def\eeqn{\end{equation}}

\def\beqa{\begin{eqnarray}}
\def\eeqa#1{\label{#1}\end{eqnarray}}
\def\eeqan{\end{eqnarray}}

\let\bar=\overbar

\def\Dslash{\not{\hbox{\kern-4pt $D$}}}
\def\dslash{\not{\hbox{\kern-2pt $\del$}}}

\def\msb{{\bar{\ssstyle M \kern -1pt S}}}

\begin{document}
\begin{titlepage}
\pubblock

\vfill
\Title{A first look at data from the NO$\nu$A upward-going muon trigger}
\vfill
\Author{R. Mina$^1$, E. Culbertson$^1$, M. J. Frank$^1$, R. C. Group$^1$, A. Norman$^2$, and\\ I. Oksuzian$^1$}
\Address{$^1$\uva}
\Address{$^2$\fnal}
\ead{ram2aq@virginia.edu}
\vfill
\begin{Abstract}
The NO$\nu$A collaboration has constructed a 14,000 ton, fine-grained, low-Z, total absorption tracking calorimeter at an off-axis angle to an upgraded NuMI neutrino beam. This detector, with its excellent granularity and energy resolution and relatively low-energy neutrino thresholds, was designed to observe electron neutrino appearance in a muon neutrino beam, but it also has unique capabilities suitable for more exotic efforts. In fact, if sufficient cosmic ray background rejection can be demonstrated, NO$\nu$A will be capable of a competitive indirect dark matter search for low-mass Weakly-Interacting Massive Particles (WIMPs). The cosmic ray muon rate at the NO$\nu$A far detector is approximately 100 kHz and provides the primary challenge for triggering and optimizing such a search analysis. The status of the NO$\nu$A upward-going muon trigger and a first look at the triggered sample is presented.
\end{Abstract}
\vfill
\begin{Presented}
DPF 2015\\
The Meeting of the American Physical Society\\
Division of Particles and Fields\\
Ann Arbor, Michigan, August 4--8, 2015\\
\end{Presented}
\vfill
\end{titlepage}

\section{Introduction}
WIMPs captured by the gravitational field of the Sun that are slowed
through collisions with solar matter can accumulate in the solar core.
There, WIMP annihilation may produce neutrinos with much larger energy
than solar neutrinos.  The signal would be an excess of high-energy
($>0.5$\,GeV) neutrino events pointing back to the Sun~\cite{Hagelin:1986gv,Buckley:2013bha}.  The cleanest
signature at NO$\nu$A will be from $\nu _{\mu}$ charged-current scattering (CC) events producing
upward-going muons that can be reconstructed in the NO$\nu$A detector.
The large and unique NO$\nu$A far detector, with its excellent granularity
and energy resolution, and relatively low-energy neutrino thresholds,
is an ideal tool for these indirect dark matter searches.

Only the upward-going flux will be considered in order to suppress
the cosmic ray-induced muon background.  The downward-going muon rate in the NO$\nu$A
far detector is approximately 100,000 Hz.

The neutrino flux from dark matter annihilation is model
dependent; however, energies from $\sim$0.5\,GeV to many TeV should
be detected with high acceptance.  For high-mass signal hypotheses,
NO$\nu$A will not be able to compete with the high acceptance of the IceCube
detector~\cite{Aartsen:2012kia}.  For lower-mass scenarios (below
$\sim$20 GeV) the Super-Kamiokande experiment currently has the
best sensitivity~\cite{Tanaka:2011uf,Choi:2015ara}.  If an efficient upward-going
muon trigger and sufficient cosmic ray background rejection can be
achieved, NO$\nu$A will be competitive with Super--Kamiokande
for WIMP mass hypotheses below 20 GeV/c$^2$.

Neutrinos are produced by interactions of high-energy cosmic rays in the atmosphere, 
and these atmospheric neutrinos comprise the primary component of the upward-going $\nu _{\mu}$ flux 
at the NO$\nu$A far detector. It is well known that such neutrinos undergo
oscillations as they travel through the Earth~\cite{Fukuda:1998ah,Fukuda:1999pp}.
Therefore, isolating the upward-going muon signal should allow for an atmospheric
neutrino oscillation study at NO$\nu$A. Since atmospheric neutrinos represent a background 
in the dark matter annihilation search, such a study is a natural preliminary step.

At NO$\nu$A, the neutrino analyses simply store events synchronous
with the NuMI beam.  For non-beam exotic physics searches, so-called
data-driven triggers~\cite{Fischler:2012zz} are required to select events of interest. 

Two data-driven triggers were developed for use at the NO$\nu$A
far detector: one that searches for long muon tracks originating from
outside the detector, and another that searches for shorter tracks
that are fully contained in the detector volume. Both triggers use
timing information for detector cell hits along the length of the track
to determine its direction. The effectiveness of the trigger at differentiating
upward- from downward-going muons in simulated events has been 
demonstrated~\cite{Mina:2015efa}.

This note will examine data from the first trigger which searches for
through-going muon tracks, and will demonstrate the effectiveness of
combining hit timing with event geometry information to isolate a small,
likely signal-rich component of the triggered sample. Producing such a
subsample is a necessary preliminary step to making an atmospheric neutrino
oscillation measurement and to performing the indirect dark matter search.

\section{Triggered Sample}
The upward-going muon trigger was first implemented and tested in August 2014,
but did not run in a stable configuration until December 2014. The triggered sample
examined in this note covers a period of 164 days from December 2014 to May 2015. 
The total livetime of this sample is $\sim$84 days.

Over the period of this sample, the through-going trigger fired at a consistent rate of $\sim$1 Hz.
Each triggered event is 50 $\mu$s in length, so that the triggered sample
contains approximately 1 part in 20,000 of the total background activity during the exposure time.
Activity in the NO$\nu$A far detector is dominated by muons from cosmic ray interactions above and 
around the detector~\cite{Mina:2015efa}.

NO$\nu$A reconstruction software was run on the triggered sample to produce the
desired track and hit objects and to perform the necessary timing calibrations.
The reconstructed sample contained 4.3\e{6} track objects.

\section{Hit Timing and LLR}
NO$\nu$A's cm-scale spatial resolution allows determination of particle direction
by comparing timing for detector cell hits along a track. By applying several
timing calibration techniques, single hit time resolution for the Far Detector
has been improved to $\sim$10 ns~\cite{Niner:2015aya}. This timing resolution is sufficient
to allow effective directionality determination using a timing-based classifier
called LLR~\cite{Mina:2015efa}.

Cleanup requirements including track length, track linearity, and number of hits
in the track object improve the reliability of the LLR as a discriminator,
and are used in the trigger to improve the determination of directionality.
Applying those requirements used in the trigger to the reconstructed sample
with full timing calibration produced a timing-based candidate subsample of 16,000 tracks
that appear to be from upward-going muons.

\section{Event Geometry}
\begin{figure}[hb]
\centering
\includegraphics[width=4.0in]{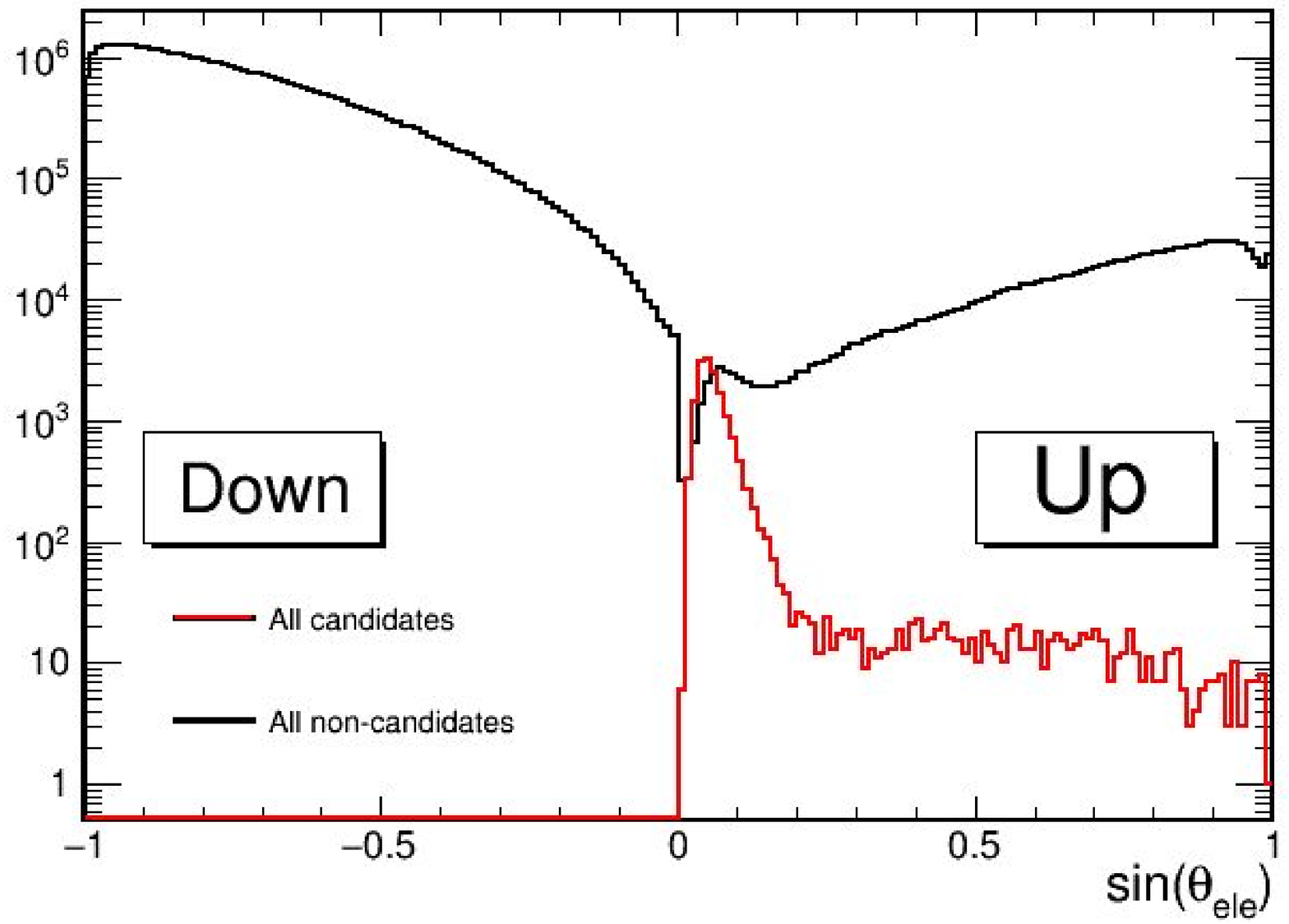}
\caption{\label{fig:EleAngle} The distribution of sine of the elevation
angle for each track in the timing-based candidate subsample (red) and all
tracks excluded from the subsample (black). Almost all candidates have
an elevation angle near 0, indicating they are nearly parallel with the ground.
A negative elevation angle indicates a downward-going track, while a positive angle
indicates an upward-going track.}
\end{figure}

The tracks passing cleanup and timing requirements are predominantly horizontal or slightly
upward-going, as shown in Fig.~\ref{fig:EleAngle}. The abundance of
mostly-horizontal tracks in this subsample is explained by the position
of the far detector on the surface, where energetic cosmic ray-induced muons travelling
slightly upward can penetrate the walls of the detector hall and
the thin layer of the Earth's crust surrounding it. At steeper angles
the horizon provides shielding from these upward-going cosmic ray muons, 
explaining the fall-off in the subsample elevation angle distribution.

\begin{figure}[hbt]
\centering
\includegraphics[width=6.0in]{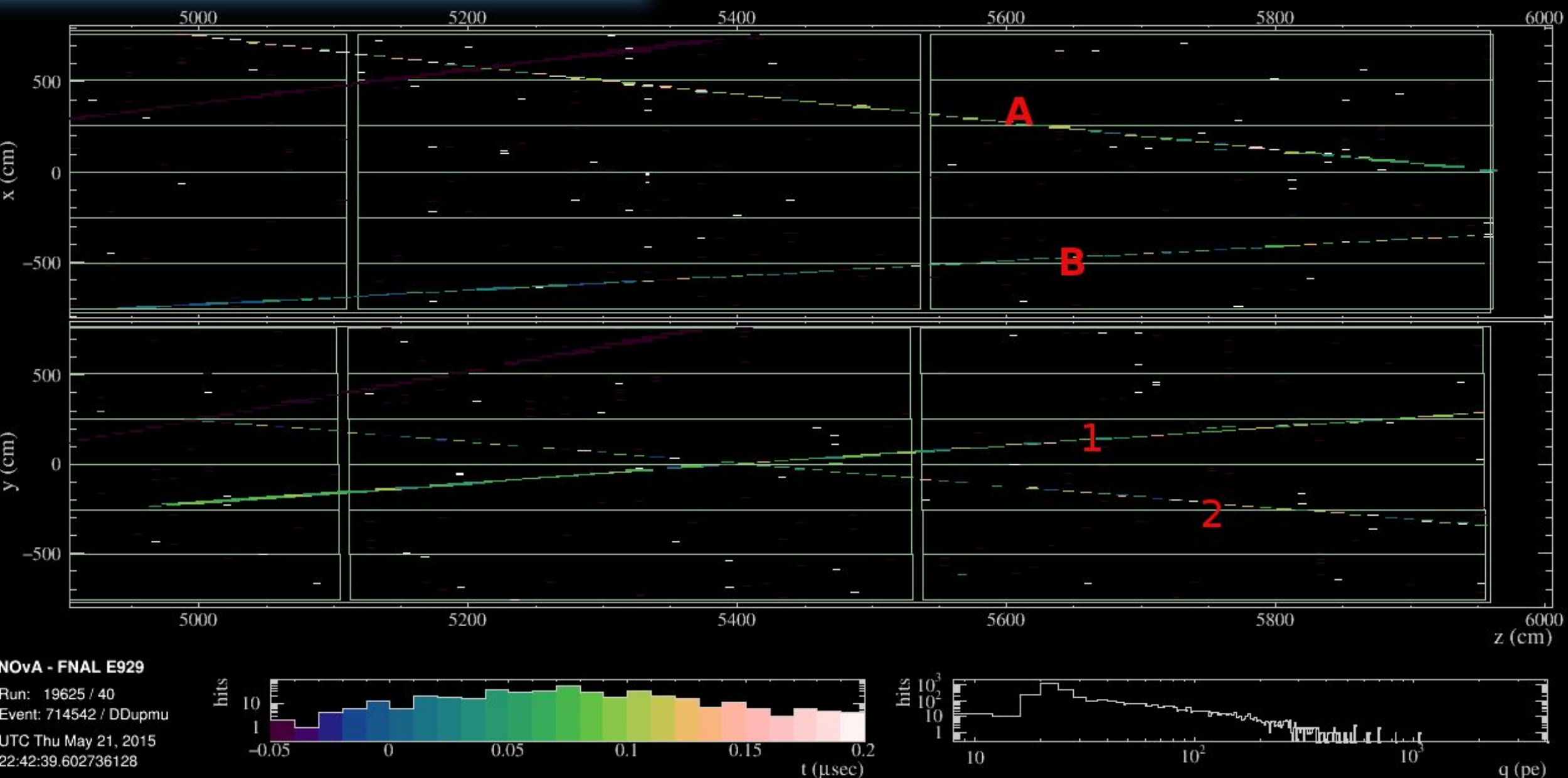}
\caption{\label{fig:Mismatch_example} An example of an event that is prone to
a possible misreconstruction. Note the two overlapping muon tracks (the two long
colored lines in each plot) with similar
extent in the $z$-dimension (horizontal axis) and near coincidence in time (hits are colored by time). The detector produces
separate two-dimensional views of each event, and the 2D track objects from each view must
be merged to produce a full 3D object. Cases such as this produce ambiguity in
matching the 2D components between the views; is the correct matching (A1,B2) or
(A2,B1)? This class of events represents the largest component of the subsample when
timing and elevation angle requirements are applied.}
\end{figure}

Placing an additional requirement on the elevation angle at 10 degrees further reduced
the size of the candidate subsample to 1,051 tracks. Of those that remain, $\sim$75\%
are not conclusively upward-going due to a possible misreconstruction in which two
unrelated but overlapping muon tracks create ambiguity in the reconstruction,
as shown in Fig~\ref{fig:Mismatch_example}. A simple geometric requirement was then applied to eliminate
this component while preserving the neutrino-induced muon signal. 255 candidates remained
after this requirement.

\section{Event Categorization}

\begin{table} [hbt]
\begin{center}
\begin{tabular}{ |l|c| } 
 \hline
 Through-going & 105 \\ \hline
 Stopping & 75 \\  \hline
 3D mismatch & 34 \\ \hline
 Up-scattered cosmics & 23 \\ \hline
 In-produced & 1 \\ \hline
 Likely downward-going & 1 \\ \hline
 Likely caused by timing miscalibration & 14 \\
 \hline
\end{tabular}
\caption{\label{tab:Categorization} Event topologies in the candidate subsample.}
\end{center}
\end{table}

Each of the remaining candidate events was then examined visually and categorized. With the
exception of 2 events that were difficult to categorize because they had attributes of multiple
event topologies, all the events could be placed into seven categories based on event topology.
The categories are summarized in Table~\ref{tab:Categorization}.

\begin{figure}[hbt]
\centering
\includegraphics[width=6.0in]{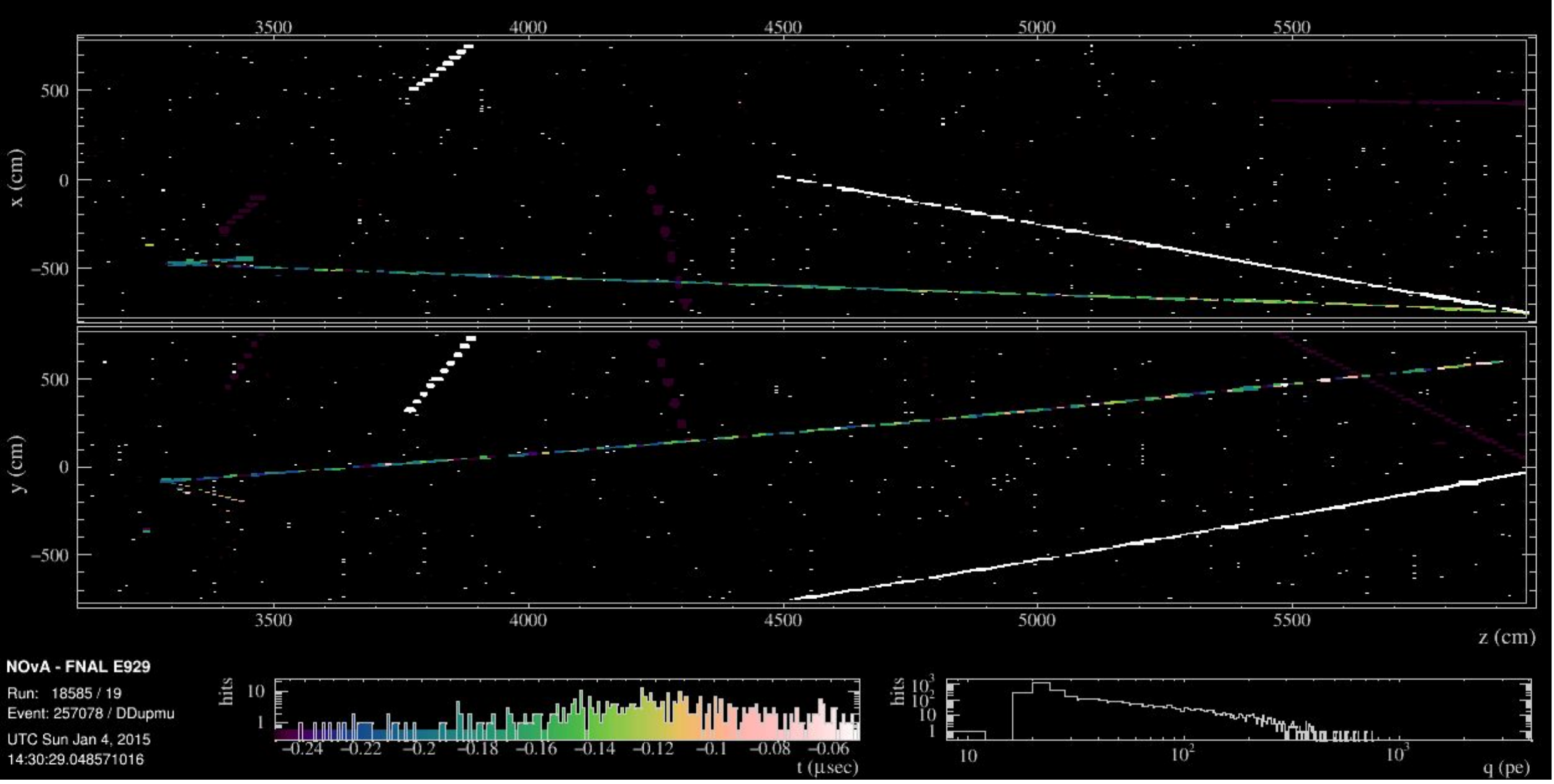}
\caption{\label{fig:In_Produced} This event contains a long upward-going muon track
that appears to have been caused by a $\nu _{\mu}$ CC interaction
within the detector.}
\end{figure}

The most common event topology was through-going tracks, indicating muons that originated
outside the detector and traveled all the way through without stopping. Both tracks
in Fig.~\ref{fig:Mismatch_example} exemplify this topology. Stopping tracks
caused by muons originating below or to the sides of the detector
that stop within the detector volume are the second most abundant
component. The final signal-like event in the subsample contains a track that appears
to be caused by a neutrino interaction within the detector. This event is shown
in Fig.~\ref{fig:In_Produced}.

\begin{figure}[hbt]
\centering
\includegraphics[width=6.0in]{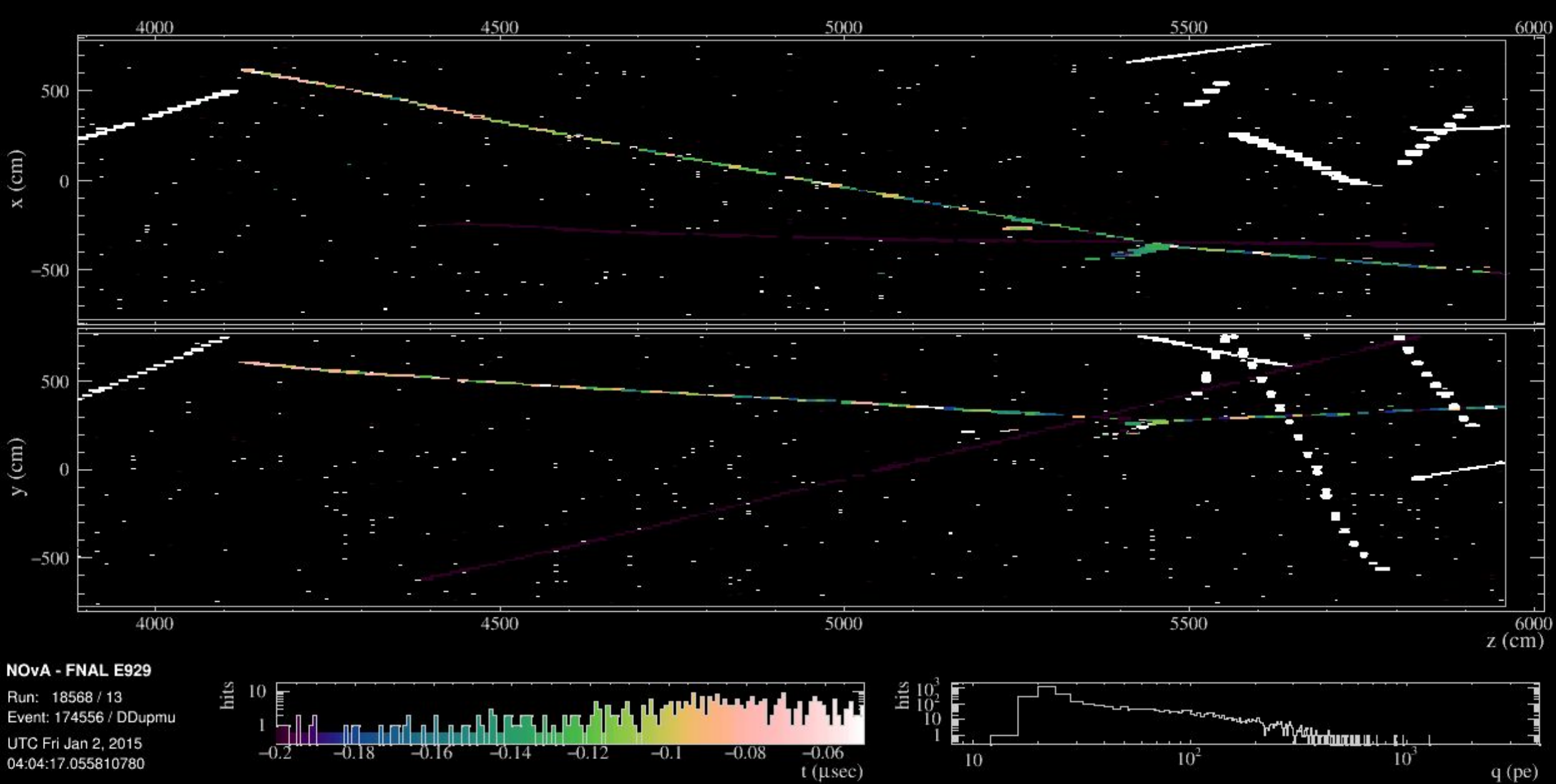}
\caption{\label{fig:Scattered} In this event, a slightly downward-going muon,
likely originating from a cosmic ray interaction outside the detector, enters from
the high-z extreme of the detector, scatters upward within it, and produces an
upward-going track.}
\end{figure}


The other categories correspond to events that are not signal-like.
34 events exemplifying the possible misreconstruction discussed in the previous section
passed the requirements (Fig.~\ref{fig:Mismatch_example}). 23 events appear to contain tracks from downward-going cosmic ray
muons that scattered upwards within the detector, as shown in Fig.~\ref{fig:Scattered}.
14 of the through-going events were placed
into a separate category because they share extraordinary features
that may indicate a temporary problem in the timing system for one portion
of the detector; namely, they all have candidate tracks that have both ends
in one particular portion of the detector, and they all occured
during an isolated, continuous running period in which the rate of through-going
events was many times higher than the average. Finally, one event appears to contain a downward-going muon.

\begin{figure}[hbt]
\centering
\includegraphics[width=4.0in]{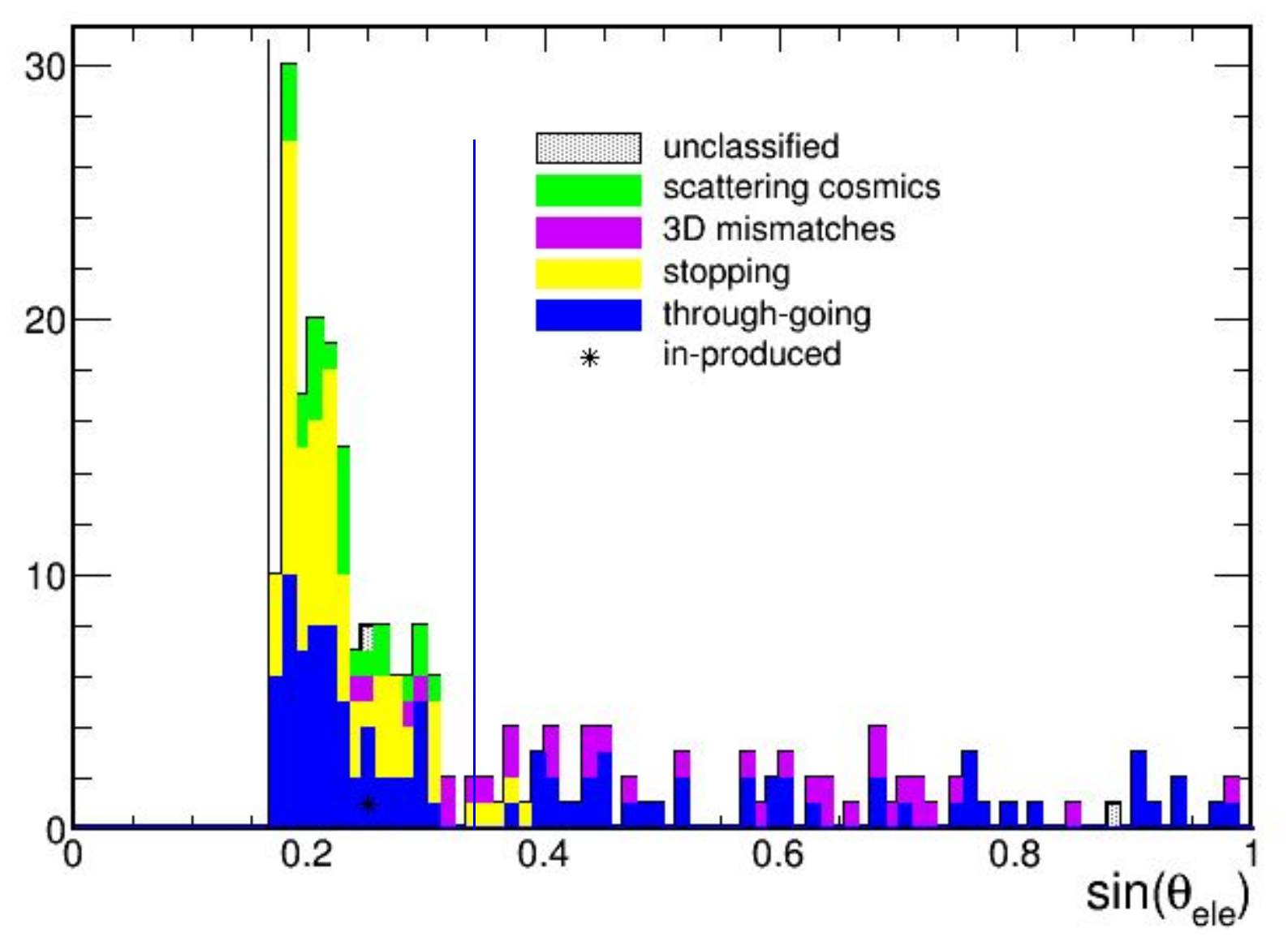}
\caption{\label{fig:Ele_categorized} The distribution of sine of the
elevation angle for some of the categorized candidate events. The black vertical
line indicates the requirement at 10 degrees. Note that above 20 degrees (the blue vertical line),
there are no events containing cosmic ray muons that scattered upward within the detector.}
\end{figure}

The existence of events that probably contain upward-going tracks from
scattering of cosmic ray muons within the detector indicates that these cosmic ray muons
are also scattering outside the detector. The through-going and stopping muon subsamples
are likely contaminated by this background process, but it is not possible
to distinguish between signal muons created by neutrino interactions 
outside the detector and those that were scattered upward in the rock around the detector.
However, all of the candidate tracks from cosmic muons scattering in the detector
had elevation angles below 20 degrees, so by requiring all tracks to have an elevation angle above 20 degrees, almost all
contamination by up-scattering cosmic ray muons should be eliminated, as shown in Fig.~\ref{fig:Ele_categorized}.
The remaining signal-like events after this requirement was applied were 43 through-going and 5 stopping muon events.

\section{Conclusions}
A first look at the triggered sample from the NO$\nu$A upward-going muon trigger
showed that in its first six months of stable running, the trigger selected
dozens of events with signal-like muon tracks. The extraction of this subsample
from the triggered sample involved a reduction by 6 orders of magnitude in the number of tracks,
and this was accomplished by combining techniques that use hit timing to determine track directionality
with simple geometry-based requirements and a visual scan. This effort revealed that
several backgrounds other than downward-going cosmic ray muons contaminate the sample,
and new requirements were developed to minimize the acceptance of these backgrounds. These techniques
will allow studies of atmospheric neutrino oscillations and, ultimately, an indirect dark matter
search at NO$\nu$A.

This work demonstrates that NO$\nu$A is capable of isolating a sample
that is likely rich in neutrino-induced upward-going muons for the through-going and
stopping muon event topologies. A similar effort that examines data from the other upward-going
muon trigger will reveal whether a signal-rich sample can be isolated for the fully-contained
event topology.

\Acknowledgments
This conference presentation was made possible by a travel award from the
American Institute of Physics Society of Physics Students. Additional
financial support was provided by the Jefferson Trust, the UVa Physics
Department, and the Fermilab Particle Physics Division. The authors
also acknowledge that support for this research was carried out by the
Fermilab scientific and technical staff. Fermilab is Operated by Fermi
Research Alliance, LLC under Contract No.~DE-AC02-07CH11359 with the
United States Department of Energy.  The University of Virginia
particle physics group is supported by DE-SC0007838.

\end{document}